\documentclass[onecolumn,amsmath,amssymb,12pt]{revtex4}

\usepackage{graphicx}
\usepackage{dcolumn}
\usepackage{bm}
\usepackage{dcolumn}

\renewcommand{\baselinestretch}{1.1}
\newcommand{\ket}[1]{\ensuremath{|#1\rangle}}

\begin{document}


\title{Macroscopic Quantum Resonance of Coupled Flux Qubits; \\\ A
Quantum Computation Scheme}

\author{Hidetoshi Akisato}
\affiliation{Department of Engineering Science, Kyoto University}
\date{\today}

\begin{abstract}
We show that a superconducting circuit containing  two loops, when
 treated with Macroscopic Quantum Coherence (MQC) theory, constitutes a
 complete two-bit quantum computer. The manipulation of the system is easily
 implemented with alternating magnetic fields. A \textit{universal} set
 of quantum gates is deemed available by means of all unitary single bit
 operations 
 and a controlled-not (\textsc{cnot}) sequence. We use multi-dimensional MQC
 theory and time-dependent first order perturbation theory to analyze
 the model. Our calculations show that a two qubit arrangement, each having a
 diameter of 200nm, operating in the flux regime can be
 operated with a static magnetic field of $\sim 0.1$T, and an
 alternating dynamic magnetic field of amplitude $\sim 1$ Gauss and
 frequency $\sim 10$Hz. The operational time $\tau_{op}$ is estimated
 to be $\sim 10$ns.
\end{abstract}

\maketitle
\begin{center}
\parbox[t]{120mm}{
{\footnotesize \tableofcontents}}
\end{center}

\newpage
\section{INTRODUCTION}
In the last few decades there has been broad interest in hope to
design and construct a practical quantum computer.  
The, so to speak, machines will enable us to reach a domain of
knowledge that was, up to now, considered unreachable or beyond human
capabilities. They will allow us to perform calculations of enormous 
amount in a rapid and effective manner. Factoring large numbers,
teleporting large amounts of information, and simulation of the
\textit{real} world, will enter our immediate reach and are destined to
dramatically changes our lives.

The core role of quantum computers are played by quantum bits,
\textit{qubits}. Many ideas have been raised for practical qubit
realization: cavity quantum electrodynamics, ion traps and nuclear spins.
Also, superconducting circuits of Josephson junctions have been proposed
and experimental research is being pursued. In this paper we will
show a possibility to implement a quantum computation scheme on a
coupled flux qubit system. In the rest of this introduction we will give
a brief description of the idea of quantum computation, and also, an outline of
the macroscopic quantum coherence theory (MQC), with which we treat our system.

\subsection{Quantum computation} \label{qcintro}
Quantum computers have attracted a lot of attention and an abundance of
work has been established. But here, we will
concentrate on only the main aspects of quantum computers. Least
requirements for a quantum computer are that the system must retain
certain properties:
\begin{enumerate}
 \item Ability to represent quantum information, meaning that a
    quantum bit must be able to represent not only two classical values
    \ket{0} and \ket{1}, but also a superposition of these two states,
    namely $\alpha\ket{0} +\beta\ket{1}$, where $\alpha$ and $\beta$ are c-numbers.

 \item Have a universal family of unitary
    transformations. \textit{Universal}, here represents that the
    Hamiltonian of the system is capable of controlling the system state
    arbitrarily. In other words one must be able to \textit{reversibly}
    transform any given state into another state of choice. The
    reversibility is a quantum mechanical requirement. DiVincenzo
    \cite{Div95} showed that a controlled-not (\textsc{cnot}) and single qubit
    gates were universal for any $n$-qubit system. 

 \item Have a preparable initial state. This is an obvious demand, since
    the initial state of the system is the input. However there is no
    necessity for capability to provide an \textit{arbitrary} initial
    state, since the above requirement suffices to transform a particular
    initial state into the desired input.

 \item Have a means of measurement. One must be able to measure the
 probability amplitude of the final state, i.e. the output. Measurement
 of the system state is
 often the result of the qubit's coupling to a classical system, commonly
 the environment. This process is
 equivalent to a projection of the superposed quantum state of the
 qubits. Although each measurement outcome is generically random, by controlling
 ensembles of quantum computers one is capable of determining the
 output state. 
 In fact, a measurement does destroy the superposition and acts randomly,
 so one must be sensitive to unwanted measurement which is a cause of
 \textit{decoherence}. 
\end{enumerate}

All quantum computers must satisfy the above four requirements, plus many
others for optimal and efficient computing. A qubit, due to its quantum
nature, suffers from decoherence, where the coherent superposition of
quantum states is destroyed by noise. This is the most influential
obstacle in creating a practical quantum computer. Extracting all noise
sources from the system is definitely impossible, but if we could
perform our operation before the system loses its coherence then we
could obtain our output with relatively high fidelity. Provided, a
strong requirement for an efficient quantum computer would be for the
`quality factor', $\tau_{d} / \tau_{op} \agt 10^4$ \cite{Orlando}. $\tau
_{op}$ is the time necessary for a single operation, and $\tau_{d}$ is
the decoherence time: the time length that the system can maintain its
coherence. 

The coupling of qubits play an important role in quantum
computation. To create large scale accurate quantum computers, there must
exist an efficient method of coupling selected qubits, and to apply
transformations upon the coupled qubits. Although coupling within the
system is permissible, often the system couples to the environment thus
resulting in considerable noise. This is an important point.

\subsection{The quantum mechanics of flux qubits}
In our research, we have chosen superconducting circuits as our qubit
arrangement. 
The advantages are that: they are relatively easy to fabricate, they
can be measured easily, and large arrays can be effectively implemented. Many of
these characteristics are results due to the fact of the flux qubit
being a macroscopic device.

However, because of its macroscopic nature an obvious question
rises: Does it behave purely quantum mechanically? In other words, can the
device be put in a \textit{superposition} of two distinct states? 
The answer to this question is provided by Macroscopic Quantum Tunneling
or Macroscopic Quantum Coherence (MQT/MQC) theory.
This field of study concerning superconducting loops was first developed
theoretically by Ivanchenko 
et. al. \cite{Ivanchenko}, and Caldeira and Leggett
\cite{Caldeira}. Experiments that followed \cite{Martinis87,Lukens95}
have confirmed so far that the magnitude of magnetic flux piercing a 
superconducting loop, when seen as a canonical variable, does obey the
rules of quantum mechanics. 

We distinguish each state of the system by
the direction of electric current, so each state is macroscopically
distinct. Here, we step out of fundamental quantum mechanics and see
that a macroscopic matter can take a state not definitely \ket{a} nor
\ket{b}, but $\alpha\ket{a} + \beta\ket{b}$. This occurs paradoxical to the
sane mind, and has been a discipline of long discussion
\cite{epr}: can the cat be dead \textit{and} alive? Although it is a
fundamental aspect, we will not indulge ourselves with a discussion of
the EPR paradox here.  

Furthermore, theory \cite{Chak83,Chak85,Lukens00c} and experiments
\cite{Lukens00a,Lukens00b} have shown that transition between 
these macroscopically distinct states can be induced by photons or
dynamically alternating magnetic fields.

Consequently, the tools of the trade for quantum computers can be
provided with superconducting circuits. However there still has not been
found proof to whether MQT/MQC theory can be applied to systems
with more than one degree of freedom, each being a macroscopic value.
Experiments \cite{SGH,HLL} have shown evidence that MQT does occur in
the thermal regime, but also indicate that the interaction between degrees of
freedom, resulting in a suppression of escape rates, cannot be
ignored. In a recent experiment \cite{ShaoLi}, the quantum regime MQT
theory seems to agree well with the
behavior of a system with more than one degree of freedom.

\section{DOUBLE QUBIT SYSTEM}
\subsection{Description of model}
\vspace{-5mm}
\begin{figure}[h]
 \includegraphics{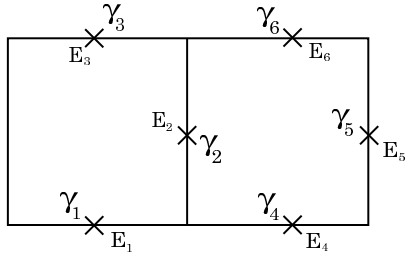}
 \caption{\label{doublequbit} The double qubit model. The crosses represent Josephson
 junctions, and the lines superconducting nodes. Each loop represents a
 qubit, and tunneling current circulates the circuit.}
\end{figure}
In this paper, we have taken a superconducting circuit as is shown in
Fig.~\ref{doublequbit}. The model consists of six Josephson junctions and
five superconducting 
nodes. The two loops of the circuit correspond to the two qubits.

\subsection{Static properties}
We assume that the temperature can be lowered well below $\Delta_G / k_B$,
where $\Delta_G$ is the gap energy of the superconductor and $k_B$ the
Boltzmann constant. Allowing this
assumption we proceed with our analysis on the basis that each node is
in a coherent 
state, i.e. maintaining a single value order parameter throughout the node.

We apply a uniform static magnetic field perpendicular to the plane the
circuit rests in. By taking the phase difference of each junction
$\gamma_i$ as our macroscopic
variable, from Josephson's Law, the tunnel current flowing through each
junction becomes,
\begin{equation} \label{tunnelcurrent}
 I_k = I_c^{\small{k}}\sin\gamma_k,
\end{equation}
where, $I_c^{\small{k}}$ is the critical current of the $k$-th junction.
From the single valuedness of the order parameter, we see that the
$\gamma$s must always satisfy the conditions,
\begin{subequations}\label{quantization}
\begin{eqnarray}
  \gamma_1 + \gamma_2 - \gamma_3 &=& 2\pi f, \\
  -\gamma_2 + \gamma_4 + \gamma_5 - \gamma_6 &=& 2\pi f,
\end{eqnarray}
\end{subequations}
where $f$ represents the magnetic flux through the loops, measured in
units of flux quanta $\Phi_0$, and is often referred to as the
\textit{frustration index}. The junction energy of this system can be
seen to be
\begin{eqnarray}
 \mathcal{U}(\vec{\gamma}) &=& \mathcal{U}_J(\vec{\gamma}) +
 \mathcal{U}_{mag}(\vec{\gamma}) \nonumber \\ 
 &=& \sum_{i=1}^{6}E_i\left(1 - \cos\gamma_i\right) +
 \mathcal{U}_{mag}(\vec{\gamma}),  
\end{eqnarray}
where $E_i$ represents the junction of energy of the $i$-th junction, and
$\mathcal{U}_{mag}(\vec{\gamma})$ is the magnetic energy of the system.
This is an analogous extension of the washboard model to
multi-dimensions. According to MQT/MQC theory, the whole system can be
regarded as a free particle with position $\vec{\gamma}$ moving within a
potential $\mathcal{U}$.

For the kinetic energy of the system we take in the electrostatic energy
of the junctions. According to Josephson's law, in a non-zero voltage
state, the voltage across the junction can be expressed as
\begin{equation}
 V_i = \frac{\hbar}{2e}\dot{\gamma_i}.
\end{equation}
Each junction has a capacitance designated as $C_i$. Provided, the
electrostatic energy of the $i$-th junction becomes
\begin{equation}
 \mathcal{T}_i(\dot{\gamma_i}) = \frac{1}{2}C_iV_i^2 
               = \frac{C_i}{2}\left(\frac{\hbar}{2e}\right)^2\dot{\gamma_i}^2
\end{equation},
and the total electrostatic energy is simply the sum, 
\begin{equation}
\mathcal{T}(\dot{\vec{\gamma}}) = \sum^{6}_{i=1}\mathcal{T}_i(\dot{\gamma_i})
\end{equation}.
In a straightforward manner, we have the Lagrangian:
\begin{eqnarray}
 \mathcal{L}\left( \dot{\vec{\gamma}},\vec{\gamma} \right) 
&=& \mathcal{T}(\dot{\vec{\gamma}}) -
 \mathcal{U}(\vec{\gamma}).
\end{eqnarray}
Here, we apply the constraints Eq.(\ref{quantization}) by letting
\begin{subequations}\begin{eqnarray}
 \dot{\gamma_1} + \dot{\gamma_2} - \dot{\gamma_3} &=& 0, \\
-\dot{\gamma_2} + \dot{\gamma_4} + \dot{\gamma_5} - \dot{\gamma_6} &=& 0,
\end{eqnarray}\end{subequations}
and the Lagrangian then becomes
\begin{eqnarray}
\mathcal{L} &=&
 \frac{1}{2}\dot{\vec{\gamma}}^{\prime t}\mathrm{C}\dot{\vec{\gamma}}^{\prime}
 - \mathcal{U}\left(\gamma^\prime\right)
\end{eqnarray}
where, $\vec{\gamma}^\prime = \left(\gamma_2, \gamma_3, \gamma_4,
\gamma_5\right)$, and $\mathrm{C}$ is the coefficient matrix of the
bilinear form for the charging energy.
Therefore the conjugate momentum of $\vec{\gamma}$ becomes
\begin{equation}
 \vec{p} =
 \frac{\partial\mathcal{L}}{\partial\dot{\vec{\gamma}}^{\prime}} =
 \mathrm{C}\dot{\vec{\gamma}}^{\prime},
\end{equation}
where $\mathrm{C}$ is a tensor of rank two. By applying a Legendre
transformation, we obtain the Hamiltonian of the system,
\begin{eqnarray}
 \mathcal{H} &=& \vec{p}^{\hspace{1mm}t}\dot{\vec{\gamma}}^{\prime} - \mathcal{L}
  \nonumber \\ 
&=& \frac{1}{2} \vec{p}^{\hspace{1mm} t} \mathrm{C}^{-1} \vec{p}^{\hspace{1mm}}
 + \mathcal{U}\left(\vec{\gamma}^{\prime}\right).
\end{eqnarray}
Notice that here, $\mathrm{C}$ is a tensor relevant to the inverse mass of a certain 
``\textit{particle}''. From now on the system under consideration will
be, given the theoretical analogy, referred to as the
``\textit{particle}'' moving within a potential
$\mathcal{U}\left(\vec{\gamma}^\prime\right)$ 

For a quantum mechanical treatment, one now introduces the
fundamental replacement,
\begin{equation} 
\hat{p} = -i\hbar\frac{\partial}{\partial\gamma}.
\end{equation}
After this replacement
the quantum mechanical Hamiltonian can be written as,
\begin{eqnarray} \label{hamil}
\mathcal{H} &=&
 \frac{1}{2} \vec{p}^{\hspace{1mm}t} \mathrm{C}^{-1} \vec{p}^{\hspace{1mm}}
 \nonumber \\[-2mm]
&& + E_1\{1 - \cos (- \gamma_2 + \gamma_3 + 2\pi f)\}
+ \sum_{i=2}^{5}E_i\left(1 - \cos\gamma_i \right) \nonumber \\
&& + E_6\{1 - \cos (- \gamma_2 + \gamma_4 + \gamma_5 - 2\pi f) \} \\[2mm]
&\equiv& \tilde{\mathcal{T}}(\gamma_2,\cdots, \gamma_5, f)
       + \tilde{\mathcal{U}}(\gamma_2,\cdots, \gamma_5, f)
\end{eqnarray}

\begin{figure}[h]
 \includegraphics{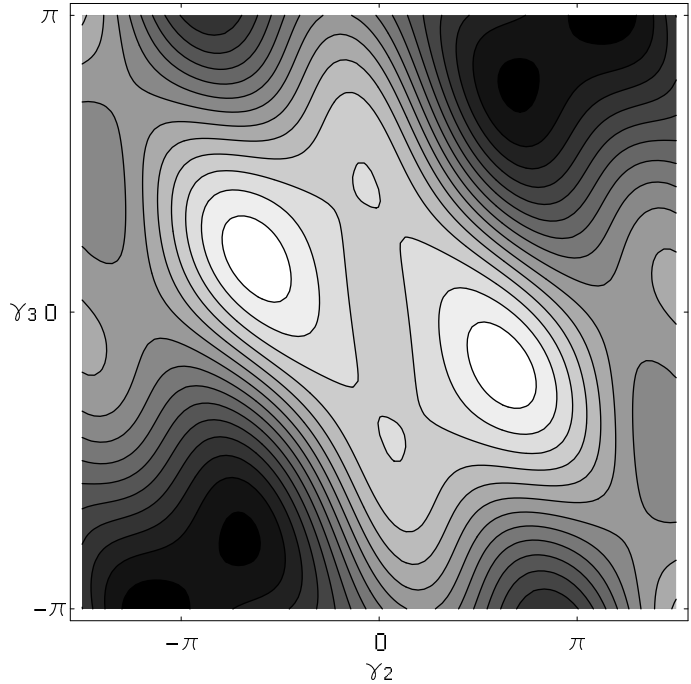}
 \caption{ \label{pot1}The reduced potential $\tilde{\mathcal{U}}(\gamma_2,\cdots,
\gamma_5, f = 0.5)$ projected on to a two dimensional space, in which the four
 local minimums (light-colored sections) of $\tilde{\mathcal{U}}$
 exists. Each local minimum
 corresponds to a distinct macroscopic state of the system, from the
 upper most in clockwise order: \ket{0}\ket{0}, \ket{1}\ket{0},
 \ket{1}\ket{1}, \ket{0}\ket{1}.}
\end{figure}

\begin{figure}[h!]
 \includegraphics{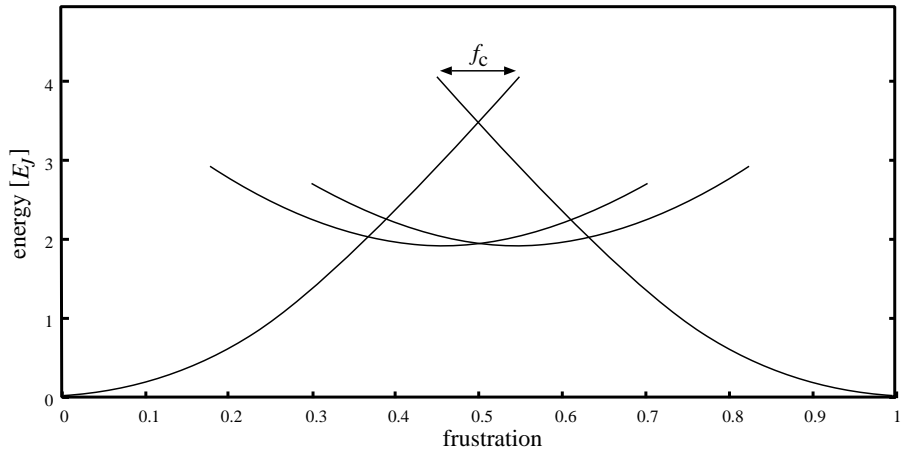}
 \caption{ \label{spectra}The energy levels of metastable states versus the frustration
 index $f$. There exits four metastable states when $0.5- f_c \le f \le
 0.5 + f_c$, $f_c \simeq 0.05$. All Josephson junction energies are set
 equal to $E_J$.} 
\end{figure}

\begin{figure}[h]
 \includegraphics{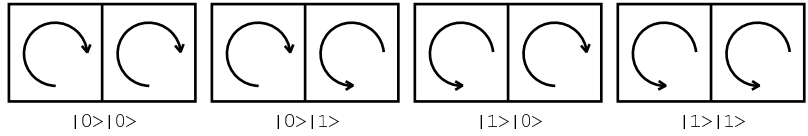}
 \caption{ \label{current}Symbolic representation of the distinct states. The arrows
 indicate the direction of the tunneling current.}
\end{figure}

To determine the motion of the ``\textit{particle}'', we study the
reduced potential term $\tilde{\mathcal{U}}(\gamma_2,\cdots, \gamma_4, f)$.
In Fig.~\ref{pot1}, we show $\tilde{\mathcal{U}}(\gamma_2,\cdots,
\gamma_5, f)$ at $f=0.5$, projected on to a two dimensional space. It is
merely a guide to
the eye, but clearly shows four metastable states, where the
``\textit{particle}'' will be able to rest.
We used a simple conjugate gradient method to
determine the local
minimums of $\tilde{\mathcal{U}}$. 
The energy level of each metastable state can be seen in
Fig.~\ref{spectra}.
One can see that, even after a slight shift in $f$, as long
as $ 0.5- f_c \le f \le 0.5 + f_c$, the system retains its four
metastable states. Calculation shows that $f_c \simeq 0.05$. 
Each of the four states differ in current direction, and the behavior
is symbolically expressed in Fig.~\ref{current}, and the numerical data
is listed in Table.~\ref{currentdata}.
\renewcommand{\baselinestretch}{1.0}
\begin{table}[h]
\caption{\label{currentdata} Values of the phase difference at each
 junction for each state, in radians. The symbol within the parentheses represent the
 tunnel current direction (through Eq.(\ref{tunnelcurrent})), in accordance with Fig.~\ref{doublequbit}.}
\begin{ruledtabular}
 \begin{tabular}[t]{ccccc} 
  & \ket{0}\ket{0} & \ket{0}\ket{1} & \ket{1}\ket{0}  & \ket{1}\ket{1} \\[1mm] \hline
Junction 1&-1.52 ($\gets$)& -0.57 ($\gets$)&0.57 ($\to$)
  & 1.51 ($\to$) \\
Junction 2 & 0.0 (-)&4.28 ($\downarrow$) &2.01 ($\uparrow$) &0.0(-) \\
Junction 3 & 1.51($\to$)& 0.57 ($\to$)& -0.57 ($\gets$) & -1.52 ($\gets$)\\
Junction 4 & -1.08 ($\gets$) & 0.38 ($\to$)& -0.38 ($\gets$) & 1.08 ($\to$)\\
Junction 5 & -1.08 ($\downarrow$) & 0.38 ($\uparrow$) & -0.38 ($\downarrow$) & 1.08 ($\uparrow$) \\
Junction 6 & 1.08 ($\to$) & -0.38 ($\gets$) & 0.38 ($\to$)& -1.08  ($\gets$) \\
 \end{tabular}
 \end{ruledtabular}
\end{table}
\renewcommand{\baselinestretch}{1.1}

Flux qubits represent each of the two states, $\ket{0}$ and
$\ket{1}$ by current direction. $\ket{0}$ corresponding to clockwise
current in the loop and $\ket{1}$, counter-clockwise. The left and right
loops will suffice to represent two distinct states, and
we can see the system is capable of providing four quantum states, 
$\ket{0}\ket{0}, \ket{0}\ket{1}, \ket{1}\ket{0}, \ket{1}\ket{1}$, which
are the basis of a two-bit quantum computer.

The important aspect of the circuit is that it contains many parameters
that can be selected by the operator. Changes in junction energies $E_i$, and
the junction capacitance $C_i$, varies the energy level of the
metastable states. Hence $f_c$ varies along, and for some parameter
settings, the system loses its four state configuration. But by
adjusting the $E_i$s and the $C_i$s, the operator is capable of preparing
optimum configuration, in accordance with the experimental factors.

\subsection{Manipulation of qubits}
The dynamical control of the qubits' state, is essential to effective
quantum computation. We now introduce a scheme for manipulating the
qubits in a simple yet efficient manner.

We set $f$ slightly away from $0.5$, yet within the four-state regime.
By this, we see from Fig.~\ref{spectra} that the four states each correspond
to local potential minimum with different energies. 
We approximate the bottom of each local minimum of $\tilde{\mathcal{U}}$
with a multi-dimensional harmonic oscillator. Hence the
``\textit{particle}'' wave function $\ket{i}\ket{j}$ will practically be a
Gaussian wave packet standing at the corresponding minimum.
Here and on, notation \ket{i,j} may be used for \ket{i}\ket{j}, and
$\epsilon_{ij}$ for its ground state energy.
We induce transitions between states by applying a time-dependent
perturbation. Physically, a time dependent magnetic field, oscillating at
certain frequencies, applied perpendicular to the circuit will result as a
perturbation fulfilling our need. We assume that the particle state ket
\ket{\psi} will always stay within the Hilbert space spanned by
{\ket{i,j}} ($i,j = 1,2$), and with this as base kets
\footnote{
Mapping from the continuous wave function problem to the
four-state problem is not trivial. For example, see
Ref.~\onlinecite{Chak85}.
}
, the Hamiltonian (\ref{hamil}) plus a time dependent perturbation
can be represented as,
\begin{eqnarray}
\label{hamil-2}
 \mathcal{H} &\equiv& \mathcal{H}_0 + \mathcal{V}\cos\left(\omega t +
 \varphi\right) \nonumber \\ 
&\doteq& \left(
\vspace{2mm}
\begin{array}{cccc}
\vspace{2mm}\epsilon_{00} & \Delta^{01}_{00}&\Delta^{10}_{00} & 0 \\
\vspace{2mm}\Delta^{01}_{00}  & \epsilon_{01}& 0 & \Delta^{11}_{01} \\
\vspace{2mm}\Delta^{10}_{00}  & 0 & \epsilon_{10}& \Delta^{11}_{10}\\
0  & \Delta^{11}_{01} & \Delta^{11}_{10}& \epsilon_{11}
\end{array}
\right) + \left(
\begin{array}{cccc}
\begin{array}{cc}
\vspace{2mm}\mathcal{V}_{00}  & \\
\vspace{2mm}  & \mathcal{V}_{01}
\end{array} 
 & 0 \\
0 & 
\begin{array}{cc}
\vspace{2mm}\mathcal{V}_{10}& \\
\vspace{2mm}& \mathcal{V}_{11} 
\end{array} 
\end{array} 
\right)\cos\left(\omega t + \varphi\right).
\end{eqnarray}
$\Delta_{ij}^{kl}$ represents the tunneling probability for transition
$\ket{i,j} \longleftrightarrow \ket{k,l}$. We have used, pfrom the
Hermiticity of the Hamiltonian, $\Delta^{kl}_{ij} = \Delta^{ij}_{kl}$.
Notice that we have introduced an approximation that, tunneling: 
$\ket{0,0} \longleftrightarrow \ket{1,1}$ and $\ket{1,0}
\longleftrightarrow \ket{0,1}$ are disallowed. Though this may lack
rigor, by looking at Fig.~\ref{pot1}, it should be clearly acceptable.
The perturbation terms,
\begin{eqnarray}\label{Velement}
 \mathcal{V}_{ij} &=& \mathcal{V}_{ij}^M + \mathcal{V}_{ij}^J \nonumber \\
 &\equiv&  \left(\sum_{k=1,2}\frac{\phi^k_{ij}}{L_k}\right)\delta\phi_{ext} 
 + 2\pi\left\{ \right. -E_1\sin\left(-\gamma_2^{ij} + \gamma_3^{ij} + 2\pi f\right) \nonumber \\
  && \hspace{35mm} + \left. E_6\sin\left(-\gamma_2^{ij} + \gamma_4^{ij} + \gamma_5^{ij} - 2\pi f \right)\right\}\delta f,
\end{eqnarray}
where, $L_k$ is the self-inductance of the $k$-th qubit, $\delta\phi_{ext}$ the
amplitude of the alternating magnetic field, $\delta f = \delta\phi_{ext}
/ \mathrm{\Phi_0}$, and 
$\phi_{ij}^k$ and $\gamma_k^{ij}$ are the magnetic flux piercing the
$k$-th qubit and the phase value, respectively, while the
system is in state $\ket{\psi_{ij}}$. $\mathcal{V}_{ij}^M$ resembles the
magnetic response of the system and $\mathcal{V}_{ij}^J$ is the response
of the Josephson junctions to the magnetic perturbation.

\begin{figure}
 \includegraphics{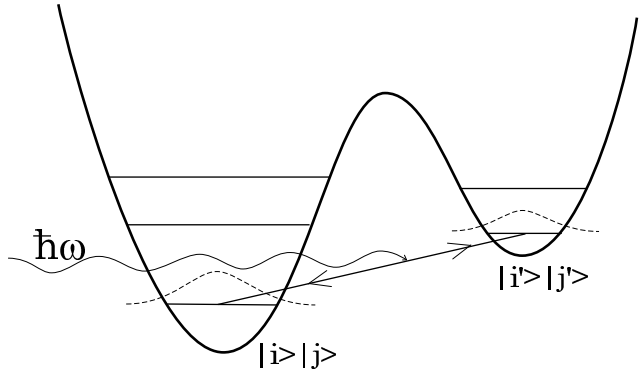}
 \caption{\label{trans}A Schematic diagram showing the transition
 process between two wave packets.}
\end{figure}

Under the approximation that tunneling probability is small, we expand
the energy eigen kets in powers of the $\Delta$s. The zeroth order term
of the Hamiltonian (\ref{hamil-2}) contains no off diagonal elements, and 
therefore no tunneling occurs. By taking in up to first order terms, the
eigen energy functions, up to a normalization factor, become
\begin{subequations}
\begin{eqnarray}
 \ket{\psi_{00}} &=& \left(2\epsilon_{00}-\epsilon_{01}-\epsilon_{10}\right)
\vert 00 \rangle +\Delta^{10}_{00}\vert 10 \rangle +
 \Delta^{01}_{00}\vert 01 \rangle, \\
 \ket{\psi_{01}} &=& \left(2\epsilon_{01}-\epsilon_{00}-\epsilon_{11}\right)
\vert 01 \rangle + \Delta^{00}_{01}\vert 00 \rangle + 
 \Delta^{11}_{01}\vert 11 \rangle, \\
 \ket{\psi_{10}} &=& \left(2\epsilon_{10}-\epsilon_{00}-\epsilon_{11}\right)
\vert 10 \rangle + \Delta^{00}_{10} \vert 00 \rangle + 
 \Delta^{11}_{10}\vert 11\rangle, \\
 \ket{\psi_{11}} &=& \left(2\epsilon_{11}-\epsilon_{01}-\epsilon_{10}\right)
\vert 11\rangle + \Delta^{01}_{11}\vert 01\rangle +
 \Delta^{10}_{11}\vert 10\rangle.
\end{eqnarray}
\end{subequations}
Off diagonal elements of the harmonic perturbation
$\mathcal{V}\cos\left(\omega t + \varphi\right)$
appears and interstate transitions are induced. For example,
\begin{subequations}
 \label{Velements}
 \begin{eqnarray}
 \langle \psi_{00}\vert \mathcal{V} \vert\psi_{01}\rangle &=& 
\left(\frac{\mathcal{V}_{00}}{2\epsilon_{01} - \epsilon_{00} -
 \epsilon_{11}} 
+ \frac{\mathcal{V}_{01}}{2\epsilon_{00} - \epsilon_{10} -
 \epsilon_{11}}
\right)\Delta^{01}_{00}, \label{Velementsa}\\
 \langle \psi_{00}\vert \mathcal{V} 
\vert\psi_{11}\rangle&=&\mathcal{O}(\Delta^2) \label{Velementsb}.
\end{eqnarray}
\end{subequations}
Eq.~(\ref{Velementsb}) is a direct consequence of the above mentioned
approximation. From Fermi's golden rule and time-dependent perturbation
theory , transitions occur only between states whose energies differ
from each other by $\hbar\omega$.  The process is described in
Fig.~\ref{trans}. 
Let $\ket{\psi\left(t\right)}$ be the particle wave function at time $t$,
and $c_{ij}(t) = \langle\psi\left(t\right)\ket{i,j}$. 
Then, the probability amplitude evolves as follows,
\begin{subequations} \label{pat}
\begin{eqnarray}
 c_{ij}\left(t\right) 
&=& 
 c_{ij}\left(0\right)\cos\left(\frac{\langle\psi_{i,j}\vert \mathcal{V}
	\ket{\psi_{k,l}}}{\hbar}t\right) 
 + ic_{kl}\left(0\right)e^{i\varphi}\sin\left(\frac{\langle\psi_{i,j}\vert
	   \mathcal{V}\ket{\psi_{k,l}}}{\hbar}t\right), \\
 c_{kl}\left(t\right) 
&=& 
 c_{kl}\left(0\right)\cos\left(\frac{\langle\psi_{i,j}\vert \mathcal{V}
	\ket{\psi_{k,l}}}{\hbar}t\right) 
 + ic_{ij}\left(0\right)e^{-i\varphi}\sin\left(\frac{\langle\psi_{i,j}\vert
	   \mathcal{V}\ket{\psi_{k,l}}}{\hbar}t\right).
\end{eqnarray}
Note that $\varphi$ is the initial phase of the harmonic perturbation.
\end{subequations}
\subsubsection{Single bit operations}\label{single}
As referred to in the introduction, single bit operations are a
necessity for quantum computation. We utilize the above Eq.(\ref{pat}) to
perform the needed operations. We regard the two states involved as an
upstate (\ket{\hspace{-1.5mm}\uparrow}) and a
downstate(\ket{\hspace{-1.5mm}\downarrow}),
and then we will be able to capitalize on our knowledge of the algebra
of the dynamics spin $\frac{1}{2}$ systems.
All unitary transformations within this two dimensinal Hilbert space can be
decomposed into a sequence two operations, $R_x\left(\theta\right)$ and
$R_y\left(\theta\right)$, which are defined as
\begin{subequations}
\begin{eqnarray}
 R_x\left(\theta\right) &\equiv& e^{-i\theta\sigma_x/2} = \left(
\begin{array}{cc}
\vspace{2mm} \cos\frac{\theta}{2} & -i\sin\frac{\theta}{2} \\
  -i\sin\frac{\theta}{2} & \cos\frac{\theta}{2}
\end{array}
\right), \\
 R_y\left(\theta\right) &\equiv& e^{-i\theta\sigma_y/2} = \left(
\begin{array}{cc}
\vspace{2mm} \cos\frac{\theta}{2} & -\sin\frac{\theta}{2} \\
  \sin\frac{\theta}{2} & \cos\frac{\theta}{2}
\end{array}
\right),
\end{eqnarray}
\end{subequations}
where $\sigma_x$ and $\sigma_y$ are Pauli matrices.
Their notation comes from the correspondence between them and rotation operations
within a three-dimensional Euclidean space.
Provided, physical implementations for $R_x\left(\theta\right)$ and
$R_y\left(\theta\right)$ (referred to as rotations from here and below)
is sufficient for realization of all single bit 
operations. With our four state system, how
do we obtain this? The solution is quite simple.

The rotation on the first bit can be decomposed into two pulses. A
rotation within the \{\ket{0,0}, \ket{1,0}\} space and a rotation within
\{\ket{0,1}, \ket{1,1}\} will rotate the first bit successfully. 
From Eq.(\ref{pat}), a pulse with frequency $\omega = \displaystyle{\frac{\epsilon_{00}
- \epsilon_{10}}{\hbar}}$ and $\varphi = -\pi$ with a duration time
$\tau =
\displaystyle{\frac{\hbar}{2\langle\psi_{0,0}\vert\mathcal{V}\ket{\psi_{1,0}}}\theta}$,
accomplishes
$R_x\left(\theta\right)$ in the first space, and another pulse with frequency $\displaystyle{\omega = 
\frac{\epsilon_{11} - \epsilon_{01}}{\hbar}}$ for time $\tau =
\displaystyle{\frac{\hbar}{2\langle\psi_{1,1}\vert\mathcal{V}\ket{\psi_{0,1}}}\theta}$
with the same $\varphi$ value will rotate the state ket in the remaining
space. $R_y\left(\theta\right)$ operations 
follows the same rule: choose the characteristic frequency, set
$\varphi = \pi/2$, and apply a pulse with the appropriate area.

\subsubsection{Multi bit operation}\label{multi}
As stated in the introduction, given a complete set of unitary
transformations for single bit operations, a \textsc{cnot}
implementation will, in general, be the last entry in the universal
set. The controlled-\textsc{not} can be expressed as, in the basis
\{\ket{0,0}, \ket{0,1}, \ket{1,0}, \ket{1,1}\},
\begin{equation}
U_{\textsc{cnot}} \doteq 
\left(
 \begin{array}{cccc}
 1 & 0 & 0 & 0 \\
 0 & 1 & 0 & 0 \\
 0 & 0 & 0 & 1 \\
 0 & 0 & 1 & 0 
 \end{array}
\right).
\end{equation}
The first bit is considered the control bit and the second the
target. The greatest advantage of this model is that its implementation
for the \textsc{cnot} gate is extremely simple. A pulse with frequency
$\omega = \displaystyle{\frac{\epsilon_{10} - \epsilon_{11}}{\hbar}}$,
$\varphi = \pi$ and a
duration of $\tau = \displaystyle{\frac{\pi\hbar}{2\langle\psi_{1,0}\vert
\mathcal{V}\ket{\psi_{1,1}}}}$ will give the desired result.\\

Sections \ref{single} and \ref{multi} show that, by using magnetic
pulses,  we have a universal set of quantum gates for our two-bit model.
Another interesting characteristic of our system is that, the
availability of Bell states (e.g. $\displaystyle{\frac{\ket{0,0} +
\ket{1,1}}{2}}$) is trivial. We expect this to become a trait of
our system when considering quantum teleportation.

\section{DISCUSSION}
Here, we will discuss the essential aspects for our model to offer
effective computation.
\subsection{Initialization}
An initial state of our model will typically be $\ket{0,0}$ or
$\ket{1,1}$. By setting the frustration index, away from the operational
point ($f \sim 0.5$), and letting it settle into its ground
state, Fig.~\ref{spectra} shows that in the regime $f \sim 0.2$ or $f
\sim 0.8$ the system has a definite stable state: $\ket{0,0} \text{ and }
\ket{1,1}$ respectively. From Sec.~\ref{qcintro} we see that our system
fulfills the initialization condition.
\subsection{Time scale}
The most intimidating enemy of quantum computers is always
\textit{decoherence}. The most subtle noise can ruin the whole attempt.
There are two important characteristic times: $\tau_{op}\text{ and
}\tau_d$ (see Sec.~\ref{qcintro}). We will give an order estimation of
our model. Let the two loops have diameter of $200$nm, and the
Josephson junctions have junction areas of $200\text{nm}^2$ by
$400\text{nm}^2$ hence junction energy $E_J\simeq 200$GHz. For the
circuit to operate in the \textit{flux} regime rather than the
\textit{charge} regime, $E_J \gg E_C = e^2/C$, which is reachable in
experiment. Let $E_C \simeq E_J/100$, the plasma frequency (eigen frequency of the
potential bottom) $\omega_p \simeq 100$GHz, self-inductance of the
circuit $L \simeq 5$pH and the tunnel current circulating the loop
$\sim1\mu$A. The energy difference of
the states would be $\sim 25$GHz. Assume that the $\Delta$s in
Eq.(\ref{hamil-2}) is around $0.1 \sim 1$GHz \footnote{To numerically evaluate the validity of this approximation,
a multi-dimensional WKB method (valley method) or a path integral
evaluation may suit, but the computational burden has put it out of
consideration.}
. By looking at Eq.(\ref{Velement}), and taking $\phi_{ij}^k \simeq
\phi_0/2 \simeq 10^{-15}\text{Tm}^2$, and applying a dynamic magnetic
field with an amplitude of $1$ Gauss, elements $\mathcal{V}_{ij}^M$
become $\sim 10^{-4}$eV, and $\mathcal{V}_{ij}^J \sim 10^{-6}$eV. 
This order estimation shows that the magnetic response
$\mathcal{V}_{ij}^M$ is the significant factor for the perturbation.
Hence for a $\textsc{cnot}$ sequence, 
$\tau_{op} \sim 10$ns. Rotation operations are of about the same or less by an order. 
To increase the decoherence time of Josephson junction circuits, many
attempts have been made \cite{Martinis02}, and coherence times of up
to $10\mu$s have been observed for single junction Josephson
qubits \cite{SiyuanHan,YangYu}. This would give us a rough estimate for
our quality factor to be $\sim 10^3$. However, we are not so optimistic since
our system includes far more junctions, hence the circuit is more
complicated. We therefore expect more noise or decoherence. Although
this does not qualify the quality factor condition (see
Sec.~\ref{qcintro}), we would not be overly surprised if experimental
development were to increase the decoherence time by an order or two.

\subsection{Comparison with other coupling methods}
Ideas for coupling multiple qubits have been raised:
placing an auxiliary superconducting loop above the circuit and
utilizing the mutual inductance between the qubit and the loop
\cite{Orlando}. Also, directly exploiting the mutual inductance of two
individual qubits is a possibility. However our model has the advantage
that because we are capable of treating the system as one element and
isolating it from the environment all together, we expect less noise to
be trapped compared to other setups. There is less chance to couple to the
environment and hence less noise.


\begin{acknowledgments}
 The author would like to thank the members of his research group for
 exhilarating discussions and stimulating inspiration,
 Prof. A. Tachibana, for his professional and 
 intuitive insights, Dr. K. Nakamura for general supervision and
 guidance, and most of all Dr. S. Tanimura, to whom all bear 
 their highest respect, for the genial support and profound knowledge he
 never comes short of bestowing. 
\end{acknowledgments}

\nocite{Nielsen}
\nocite{Berman}
\bibliography{thesis}

\end{document}